\documentclass[aps,preprint,showkeys,showpacs]{revtex4}%
\usepackage{amsfonts}
\usepackage{amsmath}
\usepackage{amssymb}
\usepackage{graphicx}%
\begin{document}
\preprint{ }
\title[ ]{Semiclassical treatment of fusion processes
\\in collisions of weakly bound nuclei}
\author{L.F.~Canto, R.~Donangelo}
\affiliation{Instituto de F\'{\i}sica, Universidade Federal do Rio de Janeiro,
C.P. 68528,
21941-972 Rio de Janeiro, Brazil}
\author{H.D.~Marta}
\affiliation{Instituto de F\'{\i}sica, Facultad de Ingenier\'{\i}a, C.C. 30,
C.P. 11000
Montevideo, Uruguay}

\begin{abstract}
We describe a semiclassical treatment of nuclear fusion reactions involving
weakly bound nuclei. In this treatment, the complete fusion probabilities are
approximated by products of two factors: a tunneling probability and the
probability that the system is in its ground state at the strong absorption
radius. We investigate the validity of the method in a schematic two-channel
application, where the channels in the continuum are represented by a single
resonant state. Comparisons with full coupled-channels calculations are
performed. The agreement between semiclassical and quantal calculations is
quite good, suggesting that the procedure may be extended to more
sophisticated discretizations of the continuum.
\end{abstract}
\maketitle

\section{Introduction}
The effects of channel coupling in fusion reactions induced by weakly bound
projectiles have attracted great interest over the last decade \cite{BCH93}.
Some theoretical studies predict strong influence of the breakup channel over
the complete fusion (CF) cross section \cite{Hu92,Ta93,DV94,Ha00,Di02,Di03}.
One of the reasons for this interest is the growing amount of experimental
data on these processes \cite{Da02,Si99,Si02,Ko98,Tr00,Al02,Re98}, which are
important both for the study of astrophysical processes as well as for the
production of superheavy nuclei.

The appropriate theoretical tool to handle this problem is the
coupled-channels method. However, its implementation becomes very complicated
for the breakup channel, which is specially important in the case of weakly
bound nuclei, since it involves an infinite number of states in the continuum.
For practical purposes, it becomes necessary to approximate the continuum by a
finite set of states, as in the Continuum Discretized Coupled-Channels method
(CDCC)\cite{NT98}. This procedure has been extended to the case of fusion
reactions in refs.~{\cite{Ha00,Di02,Di03}}. Recently, a semiclassical
alternative based on the classical trajectory approximation of Alder and
Winther (AW) \cite{AW} has been proposed \cite{AW1}. This approximation was
used to calculate breakup cross sections and the results were compared with
those of the CDCC method. The agreement between these calculations was very
good. Since this semiclassical version of the CDCC method is much simpler, it
may be a very useful tool to calculate cross sections for other channels in
reactions with weakly bound nuclei. 

Although the AW method has been extensively used for several nuclear 
reaction processes, only very recently it
was applied to the estimate of the fusion cross section \cite{Hu04}. For this
application it was considered a simplified two bound channels problem for 
which the fusion cross section obtained with the AW method was compared with 
results of a quantal coupled-channels calculation. In spite of the large 
simplification in the problem, the agreement between these two calculations 
at above-barrier energies was very good. However, the semiclassical method 
severely underestimated the fusion cross section at sub-barrier energies. 
In the present work, we also consider two channels, but instead of being
both bound, one of them represents a resonance in the continuum. We have 
assessed the validity of the semiclassical model in this case through 
comparisons with results of the corresponding full quantum mechanics 
coupled channels calculations. We have shown in particular that the good 
agreement with quantum mechanical calculations can be extended to sub-barrier 
energies through the analytical continuation of the time.

The paper is organized as follows: in section \ref{semiclassical} we review
the semiclassical treatment of nuclear reactions and give its extension to
fusion reactions. In section \ref{comparison}, we discuss the
application of this treatment to the schematic two-channel model of 
ref. \cite{Hu04}, and show that it severely underestimates the fusion 
cross section at sub-barrier energies. Then, in section \ref{improved}, we 
develop a generalization of the semiclassical model which is accurate
at sub-barrier energies and takes into account the width of the effective 
channel. The conclusions and discussion of further work are given in the 
last section.

\section{Semiclassical description of fusion}
\label{semiclassical}

As this work is devoted to reactions induced by weakly bound projectiles, the
variables employed to describe the collision are the projectile-target
separation vector, $\mathbf{r}$, and the relevant intrinsic degrees of freedom
of the projectile, $\xi$. For simplicity, we neglect the internal structure of
the target. The Hamiltonian is then given by
\begin{equation}
h=h_{0}(\xi)+V(\mathbf{r},\xi), \label{h}%
\end{equation}
where $h_{0}(\xi)$ is the intrinsic Hamiltonian and $V(\mathbf{r},\xi)$
represents the projectile-target interaction. The eigenvectors of $h_{0}(\xi
)$\ are given by the equation%
\begin{equation}
h_{0}~\left\vert \phi_{\alpha}\right\rangle =\varepsilon_{\alpha}~\left\vert
\phi_{\alpha}\right\rangle . \label{av}%
\end{equation}
The AW method \cite{AW} is implemented in two-steps. First, one employs
classical mechanics for the time evolution of the variable $\mathbf{r}$. The
ensuing trajectory depends on the collision energy, $E,$ and the angular
momentum, $\hbar l$. In its original version, an energy symmetrized Rutherford
trajectory $\mathbf{r}_{l}(t)$ was used. In our case, the trajectory is the
solution of the classical equations of motion with the potential
$V(\mathbf{r)=}$\ $\left\langle \phi_{0}\right\vert V(\mathbf{r}%
,\xi)\left\vert \phi_{0}\right\rangle ,$ where $\left\vert \phi_{0}%
\right\rangle $\ is the ground state of the projectile. In this way, the
coupling interaction becomes a time-dependent interaction in the $\xi$-space,
$V_{l}(\xi,t)\equiv V(\mathbf{r}_{l}(t),\xi).$ 
The second step consists of treating the dynamics in the intrinsic space as 
a time-dependent quantum mechanics problem. Expanding the wave function in 
the basis of intrinsic eigenstates,
\begin{equation}
\psi(\xi,t)=\sum_{\alpha}a_{\alpha}(l,t)~\phi_{\alpha}(\xi)~e^{-i\varepsilon
_{\alpha}t/\hbar}, \label{exp}%
\end{equation}
and inserting this expansion into the Schr\"{o}dinger equation for $\psi
(\xi,t),$ one obtains the AW equations%
\begin{equation}
i\hbar~\dot{a}_{\alpha}(l,t)=\sum_{\beta}~\left\langle \phi_{\alpha
}\right\vert V_{l}(\xi,t)\left\vert \phi_{\beta}\right\rangle
~e^{i\left(  \varepsilon_{\alpha}-\varepsilon_{\beta}\right)  t/\hbar
}~a_{\beta}(l,t). \label{AW}%
\end{equation}
These equations are solved with the initial conditions $a_{\alpha
}(l,t\rightarrow-\infty)=\delta_{\alpha0},$ which means that before the
collision ($t\rightarrow-\infty$) the projectile was in its ground state. The
final population of channel $\alpha$ in a collision with angular momentum $l$
is $P_{l}^{(\alpha)}=\left\vert a_{\alpha}(l,t\rightarrow+\infty)\right\vert
^{2}$ and the angle-integrated cross section is
\begin{equation}
\sigma_{\alpha}=\frac{\pi}{k^{2}}~\sum_{l}\left(  2l+1\right)  ~P_{l}%
^{(\alpha)}. \label{PW1}%
\end{equation}

To extend this method to fusion reactions, we start with the quantum
mechanical calculation of the fusion cross section in a coupled channel
problem. For simplicity, we assume that all channels are bound and have spin
zero. The fusion cross section is a sum of contributions from each channel.
Carrying out partial-wave expansions we get
\begin{equation}
\sigma_{F}=\sum_{\alpha}~\left[  \frac{\pi}{k^{2}}\sum_{l}\left(
2l+1\right)  ~P_{l}^{F}(\alpha)\right]  , \label{sigTF}%
\end{equation}
with
\begin{equation}
P_{l}^{F}(\alpha)=\frac{4k}{E}~\int dr~\left\vert u_{\alpha l}(k_{\alpha
},r)\right\vert ^{2}~W_{\alpha}^{F}(r). \label{PQM}%
\end{equation}
Above, $u_{\alpha l}(k_{\alpha},r)$ represents the radial wave function for
the $l^{th}$-partial-wave in channel $\alpha$ and $W_{\alpha}^{F}$ is the
absolute value of the imaginary part of the optical potential associated to
fusion in that channel.

\bigskip

To use the AW method to evaluate the fusion cross section, we make the
approximation
\begin{equation}
P_{l}^{F}(\alpha)\simeq\bar{P}_{l}^{\left(  \alpha\right)  }~T_{l}^{(\alpha
)}(E_{\alpha}). \label{PLAW}%
\end{equation}
In eq.(\ref{PLAW}), $\bar{P}_{l}^{\left(\alpha\right)}$ 
is the probability that the 
system is in channel $\alpha\ $ at the point of closest approach on the 
classical trajectory, and $T_{l}^{(\alpha)}(E_{\alpha})$ is the probability 
that a particle with energy $E_{\alpha}=E-\varepsilon_{\alpha}$ and reduced 
mass $\mu=M_{P}M_{T}/\left(M_{P}+M_{T}\right)$, where $M_{P}, M_{T}$ are 
respectively the masses of the projectile and target, tunnels through the 
potential barrier in channel $\alpha$.

We now proceed to study the CF cross sections in reactions
induced by weakly bound projectiles. For simplicity, we assume that the GS is
the only bound state of the projectile and that the breakup process produces 
only two fragments, $F_{1}$ and $F_{2}$. In this way, the labels $\alpha=0$ 
and $\alpha\neq0$ correspond respectively to the GS and the breakup states 
represented by two unbound fragments. Neglecting any sequential contribution, 
the CF can only arise from the elastic channel. In this way, the 
cross section $\sigma_{CF}$\ can be obtained from eq.(\ref{sigCF}), dropping 
contributions from $\alpha \ne 0$. That is,
\begin{equation}
\sigma_{CF}=\frac{\pi}{k^{2}}\sum_{l}\left(  2l+1\right)  ~P_{l}^{Surv}%
~T_{l}^{(0)}(E), \label{sigCF}%
\end{equation}
where 
\begin{equation}
P_{l}^{Surv}\equiv \bar{P}_{l}^{\left(0\right)}=\left\vert a_{0}%
(l,t_{ca})\right\vert ^{2} \label{Psurv}
\end{equation}
is usually called survival (to breakup) probability. 

\section{Comparison with coupled-channels calculations}
\label{comparison}%

The accuracy of the semiclassical fusion cross section has recently been
checked in a schematic two-channel calculation for the scattering of $^{6}$He
projectiles on a $^{238}$U target at above-barrier energies \cite{Hu04}. The
weakly bound $^{6}$He nucleus dissociates into $^{4}$He and two neutrons, with
threshold energy $B=0.975$ MeV. The elastic channel is strongly coupled to the
breakup channel and the influence of this coupling on the fusion cross section
is very important. In this calculation, the breakup channel was represented by 
a single effective state~\cite{Ca03}. For simplicity, the effective channel was
treated as a bound state but it was assumed to contribute only to incomplete
fusion.  Furthermore, the ingoing wave boundary condition is used in all these
calculations. The CF cross section is therefore given by
eq. (\ref{sigCF}). In ref. \cite{Hu04} the threshold energy was neglected and 
the same potential barrier was used for both channels, {\it i.e.} Wood-Saxon 
shapes for the real and imaginary potentials, with the parameters: 
$V_{0}=-50$ MeV, $r_{0r}=1.25$ fm, $a_{r}=0.65$ fm, $W_{0}=-50$ MeV,
$r_{0i}=1.0$ fm and $a_{i}=0.65$ fm. This calculation
adopted a form factor with the radial dependence of the electric dipole 
coupling and the strength was chosen arbitrarily, so that the coupling 
modified appreciably the cross section predicted by the one dimension barrier 
penetration model. The results (figure 1 of ref. \cite{Hu04}) are shown
here in figure 1 (a). Above the Coulomb barrier, the semiclassically
calculated cross section is in very good agreement with the one obtained by
the coupled-channels method. However, the situation is much worse at energies 
below the Coulomb barrier. This is illustrated in figure 1 (b), where the 
results of (a) are plotted on a logarithmic scale. As it usually occurs, the
total fusion cross section at sub-barrier energies is enhanced in the CC 
calculation. The contribution from CF alone (solid line) is slightly larger 
than the fusion cross section in the no-coupling limit (dashed line). It is 
clear that the semiclassical calculation does not reproduce this effect, since
its CF cross section is much lower in this energy range.  This deficiency is 
due to the fact that in the semiclassical procedure described in section 
\ref{semiclassical} the coupling is not considered inside the barrier region, 
because there is no classical trajectory connecting points inside this region. 
Therefore, the effective barrier lowering ocurring in the coupled-channels 
calculation cannot be reproduced. The semiclassical calculation shows also 
some instability for energies close to the barrier top ($V_B\simeq 20$ MeV). 
This last behavior results from orbiting effects in the classical trajectories. 
\begin{figure}
[ptb]
\begin{center}
\includegraphics[ width=8cm,angle=-90] {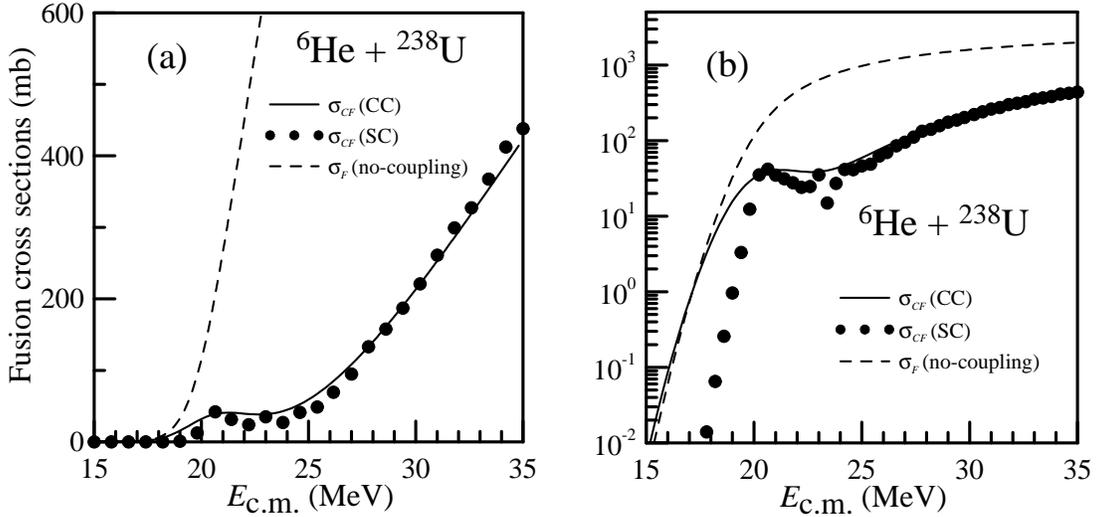}
\caption{Complete fusion cross section as calculated quantum mechanically
(full line) compared with that of the Alder and Winther calculation (full
dots) for the two channel case with $B = 0$. (See text for
details). For comparison, the fusion cross section calculated by quantum
mechanics in the no coupling case is also shown (dashed line).}\label{f1}
\end{center}
\end{figure}

\section{The improved semiclassical two-channel model}\label{improved}

\begin{figure}
[ptb]
\begin{center}
\includegraphics[ width=10cm] {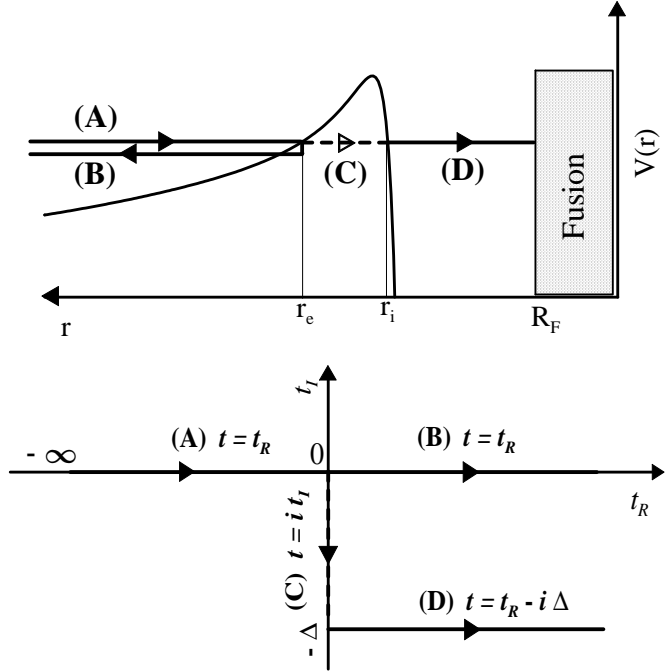}
\caption{Analytical continuation of the time variable. The upper panel shows
the branches of the classical trajectory and the lower panel the 
evolution on the complex time plane.}\label{ct}
\end{center}
\end{figure}

To improve the semiclassical model at sub-barrier energies, we resort to the 
analytical continuation method, which consists of introducing the imaginary 
part of the time variable to obtain a classical trajectory in the sub-barrier 
region \cite{LS84}. This procedure is illustrated in figure \ref{ct}, where
the time scale is chosen such that $t=0$ at the external turning point, $r_e$. 
Along the incident branch of the trajectory (A), the time develops on the 
real axis as the system approaches the barrier. At $r=r_e$, the trajectory 
splits into two parts: the reflected branch (B) and the classically forbidden 
transmission branch (C). On the former, which is not relevant to
the fusion process, the time remains real. Along the branch (C), the real 
part of $t$ remains equal to zero while its imaginary part develops on
the negative part of the imaginary axis, until this trajectory reaches the
exit point $r_i$, at $t=-i\Delta$. This trajectory is then continued 
into the internal classically allowed region (D), towards the strong 
absorption radius, $R_F$, where fusion occurs. Over this branch, the real
part of $t$ grows whereas its imaginary part keeps the value $t_I=-\Delta$.
\begin{figure}
[ptb]
\begin{center}
\includegraphics[width=8cm] {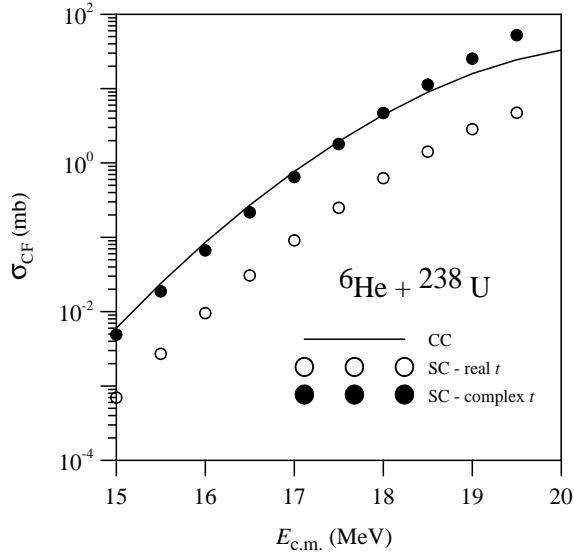}
\caption{Complete fusion cross section obtained with the calculations 
discussed in the text. }\label{Q0-AW}
\end{center}
\end{figure}

We used the analytical continuation of $t$ in the calculation of the classical 
trajectories and generalized the AW equations (\ref{AW}) accordingly.
In figure  \ref{Q0-AW}, CF cross sections of different 
calculations at sub-barrier energies are compared. The open and the full 
circles are respectively results from semiclassical calculations with real 
and complex times. The solid line represents results of coupled channel 
calculations. It is clear that the analytical continuation of the time 
variable improves substantially the accuracy of the semiclassical method. Now,
their results agree very well with the ones of the quantum mechanical 
calculation, except for the last energy point, which is close to the barrier top.
In this case, the cross section is strongly influenced by the orbiting effect
in the classical trajectory. The system remains a very long time near the 
barrier radius, where the coupling is very strong. In this way, the elastic 
channel recovers the amplitude lost to the excited channel along the trajectory 
and the CF cross section becomes very large. This is a limitation of the 
semiclassical method. However, the situation tends do be much better in a 
realistic treatment of the continuum, with many channels and continuum-continuum
coupling, owing to the irreversibility of the breakup process. The same occurs 
in the two-level model if the effective channel is treated as a doorway with 
finite width. This will be shown below.

\begin{figure}
[ptb]
\begin{center}
\includegraphics[width=10cm] {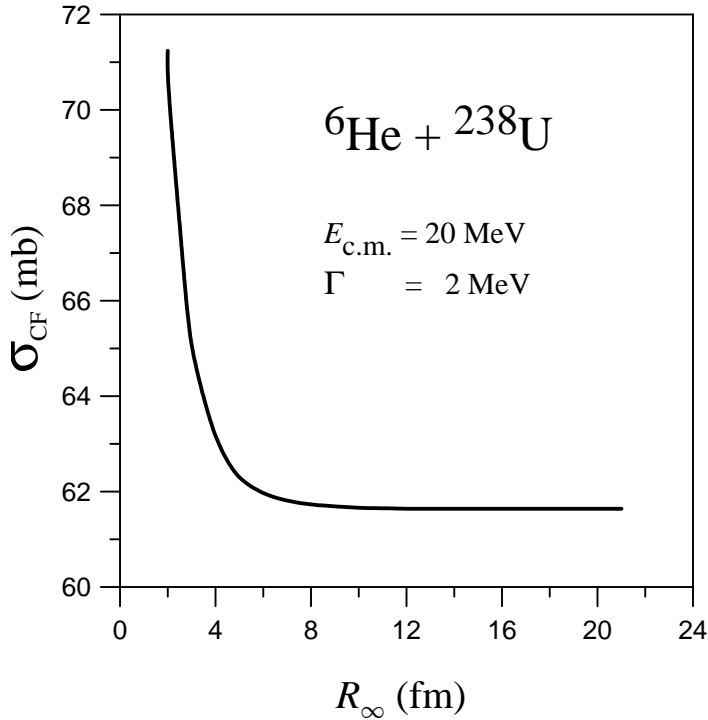}
\caption{Quantum calculation of the complete fusion cross section as a
function of the cut-off radius of the potential $-i \Gamma /2$. For details
see the text.}\label{Rinf}
\end{center}
\end{figure}

To further improve the semiclassical calculation, we take into account the 
reaction $Q$-value. We assume that the excitation energy of the effective 
channel is the breakup threshold, $B$. The next step is to simulate the 
irreversible nature of the breakup process, identified in the CDCC 
calculations of Diaz-Torres and
Thompson \cite{Di02} by an effective channel with complex energy.
To this end we use the fact that an exponentially decaying state with mean 
life $\tau = \hbar / \Gamma$, can be obtained through the inclusion of a 
constant imaginary potential equal to $-i \Gamma /2$ in the system Hamiltonian. 
This procedure, however, requires some care. The solution of the AW
equations does not present difficulties since the population of the resonant 
state is vanishingly small as $t\rightarrow -\infty$. The numerical solution 
of the coupled-channel equations, however, requires attention. To handle this
situation one should switch-off the $-i \Gamma /2$ imaginary potential at
some distance $R_\infty$ much larger than the range of the potentials, and
then match the radial wave functions with their asymptotic forms. To illustrate
this procedure, we show in figure \ref{Rinf} the CF cross section obtained
through the solution of the coupled channel equations as a function of the 
cut-off radius. For this example, we considered the collision energy 
$E_{c.m.}=20\ {\rm MeV}\simeq V_B$ and the typical width $\Gamma = 2$ MeV. 
The figure indicates that the result does not depend on the cut-off radius, 
provided that it is large enough ($R_\infty \gtrsim 20$ fm).

\begin{figure}
[ptb]
\begin{center}
\includegraphics[width=8cm,angle=-90] {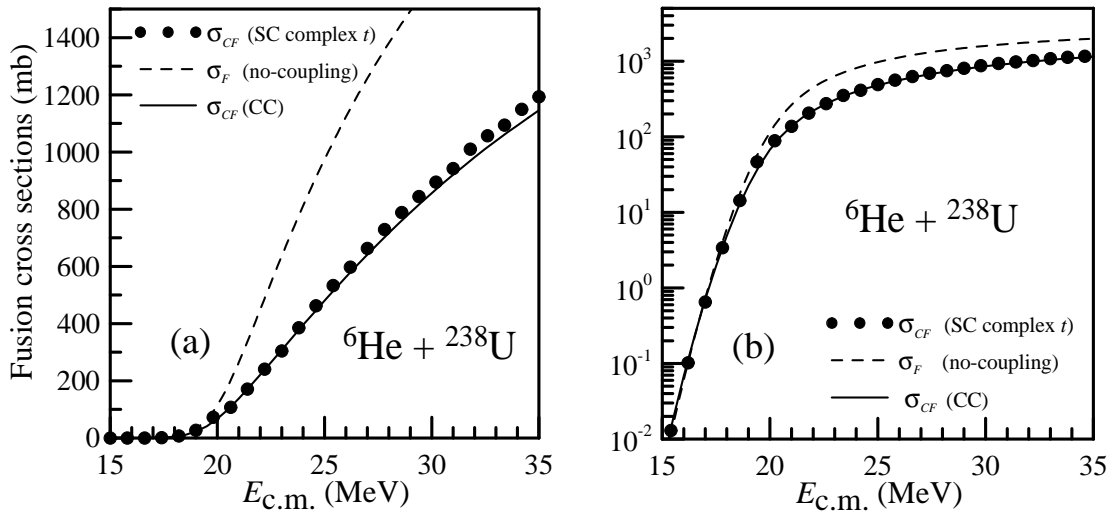}
\caption{Complete fusion cross section as calculated quantum mechanically
(full line) compared with that of the present work (full circles), including
the analytic continuation in the classically forbidden regions, for the
two channel case with $B = 0.975$~MeV and $\Gamma = 2$~MeV.}\label{CT-width}
\end{center}
\end{figure}

Let us now consider the CF cross sections obtained with the above discussed
procedures. The results of the improved semiclassical calculation (solid 
circles) are shown in figure \ref{CT-width}, in comparison with results 
of the CC method (solid line) and in the no-coupling limit. In order to
exhibit the details above and below the barrier, the cross sections are
represented on a linear (a) and on a logarithmic (b) scale. Firstly, one
observes that the suppression of the CF cross section at above-barrier 
energies is less pronounced than in figure 1. The reason for this difference
is that here we are taking into account the breakup threshold energy,
$B=0.975$ MeV, while the calculations of figure 1 were performed in the 
sudden limit ($B=0$). Nevertheless, the suppression of the CF cross section 
remains quite. Comparing the semiclassical estimate for $\sigma_{CF}$ with
the CC values, we conclude that the improved semiclassical model 
leads to accurate results, above and below the Coulomb barrier.

\bigskip
\section{Conclusions}
\bigskip We have illustrated how to extend the semiclassical method of Alder
and Winther to the case of fusion reactions. Comparison of these calculations
with full quantum ones for the case of only two channels, the elastic and an
effective one, shows a good agreement between them. Below the barrier we have
used the analytic continuation of the trajectory in the classically forbidden
region, and we have shown it essential to describe correctly the process. 

The fluctuations observed in the semiclassically calculated cross sections, 
in the case in which the effective channel has an infinite lifetime (zero
width), may be associated to orbiting processes. However, when the resonance
width is nonzero, the fluctuations vanish and the semiclassical calculations
are in very good agreement with the full quantum ones. Thus it appears that
fusion processes in weakly bound systems in which the elastic channel is
coupled to the continuum are amenable to be treated by means of the extensions
of the Alder and Winther's method presented here. 

The extension of this semiclassical method to less schematic treatments of
the continuum is presently under way~\cite{Longpaper}. In it the continuum 
is represented by a discrete set of channels, as in \cite{Di02}; therefore the relative motion between the breakup fragments is more accurately described. 
In this way, detailed information on the positions and momenta of the projectile
fragments at their creation point would be available. This would allow to 
study the collision of the fragments with the target, and therefore to
estimate the incomplete fusion cross section as well as the sequential
fusion contribution to the CF cross section. An estimate of
this last process is not possible in the standard CDCC description of 
fusion \cite{Di02}, so semiclassical calculations may furnish a way to solve 
this problem. It should be remarked that a semiclassical study along these
lines has already been implemented for calculations of the breakup cross
section \cite{AW1} and the results were in very good agreement with the
ones obtained with CDCC calculations.

\bigskip

We acknowledge useful discussions with M.S. Hussein and P.R. Silveira 
Gomes. This work was
supported in part by MCT/FINEP/CNPq(PRONEX) under contract no. 41.96.0886.00,
PROSUL and FAPERJ (Brazil), and from PEDECIBA and CSIC (Uruguay).

\end{document}